\documentclass[aps,twocolumn,nofootinbib]{revtex4}
\usepackage[latin1]{inputenc}
\usepackage{epsfig}
\usepackage{mathrsfs}
\newcommand{\beq}{\begin{equation}}
\newcommand{\eeq}{\end{equation}}
\newcommand{\la}{\langle}
\newcommand{\ra}{\rangle}

\begin{document}

\title{Classical stochastic approach to quantum
mechanics and quantum thermodynamics}

\author{Mário J. de Oliveira}
\affiliation{Universidade de São Paulo,
Instituto de Física, Rua do Matão, 1371,
05508-090 São Paulo, SP, Brazil}

\begin{abstract}

We derive the equations of quantum mechanics and quantum
thermodynamics from the assumption that a quantum system
can be described by an underlying classical system of
particles. Each component $\phi_j$ of the wave vector is
understood as a stochastic complex variable whose real
and imaginary parts are proportional to the coordinate
and momentum associated to a degree of freedom of the
underlying classical system. From the classical stochastic
equations of motion, we derive a general equation for the 
covariance matrix of the wave vector which turns out to be
of the Lindblad type. When the noise changes only the 
phase of $\phi_j$, the Schrödinger and the quantum Liouville
equation are obtained. The component $\psi_j$ of the wave
vector obeying the Schrödinger equation is related to
stochastic wave vector by $|\psi_j|^2=\la|\phi_j|^2\ra$.

\end{abstract}

\maketitle

\section{Introduction}

A distinguishing feature of quantum mechanics
\cite{landau1958,merzbacher1961,messiah1961,sakurai1967,
sakurai1994,griffiths1995,piza2002,griffiths2002}
is its formulation in terms of an unobservable: the
wave function. The presence of unobservables in a theory
does not make it unscientific, as long as they lead to
observables, that is, to quantities that can be observed
or measured experimentally \cite{oliveira2021}.
Any scientific theory has unobservables to a greater
or lesser extent. Significant examples are the epicycles
of Ptolemy, the aether of Newton, and the luminiferous
aether of Maxwell. Much less obvious examples are the
concepts of time reversibility, causality, homogeneity
of time and homogeneity of space. The unobservable of
quantum mechanics was present in its very beginning 
when Schrödinger formulated his equation in terms of
the wave function. Although the wave function is an
{\it unobservable} it leads through this equation to
the {\it observable} spectral lines of hydrogen.

The question that we address here concerns the 
possibility of the formulation of quantum mechanics
an unobservable other than the wave function.
Specifically, the unobservable that we consider here
to describe a quantum system is a system of particles
obeying the classical equations of motion
\cite{lanczos1949,goldstein1950,landau1960,arnold1978},
which we call the {\it underlying} system to avoid
confusion with a real system described by classical
mechanics.

The standard formulation of quantum mechanics postulates
that the quantum states are represented by wave vectors
belonging to a complex vector space, the Hilbert space.
In contrast, classical mechanics is represented in the
Hamilton formulation by canonical variables belonging
to a real vector space, the phase space. Thus a classical
approach to quantum mechanics, requires a formulation
of classical mechanics in terms of complex canonical
variables. 

The possibility of using complex variables to express
the classical Hamilton equations of motion was pointed
out by Lanczos \cite{lanczos1949} who showed that a
pair of complex conjugate variables is also a pair of
canonical variables. The formulation of the quantum
equation of motion by a classical Hamilton equation
in complex variables was given by Strocchi
\cite{strocchi1966}. In his formulation a complex
canonical variable is identified as a component
$\phi_j$, which obeys the complex Hamilton equation
associated to the Hamiltonian function ${\cal H}$, 
identified as the mean value
\beq
{\cal H} = \sum_j \phi_j^* H_{jk} \phi_k
\label{2}
\eeq
of the Hamilton operator $H$.

An equivalent approach was proposed by Heslot 
\cite{heslot1985} but instead of using a complex
Hamilton equation, he shows that the real and
imaginary parts of the wave functions are a pair
of real classical canonical variables obeying 
the real standard Hamilton equation. The classical
representation was then analyzed and explored by
several authors 
\cite{gray1994,bodurov1998,briggs2012,elze2012}.

A essential aspect of the classical representation
concerns the norm of the wave vector,
\beq
{\cal N} = \sum_j \phi_j^* \phi_j,
\label{2a}
\eeq
which is a quantity conserved by the Hamilton equations
of motion associated to the classical Hamiltonian
(\ref{2}). This is a nice property since according the
wave vector should be normalized at all times. However,
the conservation property does not determine the
value of the norm. Therefore we should postulate that
the norm has the same value for any possible trajectory
in phase space as this is a basic postulate of quantum
mechanics. More precisely, among all sectors of the phase
space determined by distinct values of the norm, we must
select just one of them. As we shall see, the one to be
selected is connected to the Planck constant. 

Quantum mechanics is understood as having a probabilistic
character. However this character is not clearly manifested
in the usual representation. For instance, no variable is
considered to be a random variable. The probabilistic
character is a consequence of the standard interpretation
of quantum mechanics 
\cite{omnes1994,auletta2001,freire2022}
which introduces probability in an ad-hoc and a posteriori
manner by the proposition that the square of the absolute
value of the wave function is a probability.

Here we introduce the probabilistic character in an
explicit and a priori form by turnig $\phi_j$ into a
stochastic variable. This is accomplished by transforming
the equation of motion into a stochastic equation through
the addition of a noise term in the Hamilton equations
of motion
\cite{vankampen1981,risken1989,gardiner2009,tome2015}.
The noise transforms the trajectories in the complex
phase space into stochastic trajectories, and it is set
up in such a way as to preserve the norm given by
(\ref{2a}) along any stochastic trajectory. 

The noise changes in general the absolute value and the
phase of $\phi_k$. A meaningful result of our analysis is
that the Schrödinger and the quantum Liouville equations
are obtained through noises that change the phase but not
the absolute value of $\phi_k$. This type of noise also
makes each term of the norm constant. If the noise changes
both the phase and the absolute value of $\phi_k$ then we
reach the quantum thermodynamic equation which turns out
to be the Lindblad equation for open system
\cite{breuer2002,manzano2020,lindblad1976,gorini1976}.

The quantum thermodynamic equation is the central equation
of the quantum thermodynamics that we develop here. Based
on this equation we derive the first law of thermodynamics
and the second law of thermodynamics. This is accomplished
by defining the quantum entropy and the quantum entropy
production. The second law of thermodynamics is obtained
by demonstrating that the entropy production is nonnegative.
To demonstrate this proposition we use a reasoning similar
to that employed by Spohn \cite{spohn1978} which in turn
was based on a theorem of Lieb \cite{lieb1973}.

\section{Stochastic equation of motion}

\subsection{Underlying classical system}

The underlying system consists of a classical system
with $n$ degrees of freedom. Each degree of freedom
corresponds to a pair of canonical variables $q_i$
and $p_i$, and the  Hamilton equations of motion are 
\beq
\frac{dq_i}{dt} = \frac{\partial{\cal H}}{\partial p_i},
\qquad
\frac{dp_i}{dt} = -\frac{\partial{\cal H}}{\partial q_i},
\eeq
which determine trajectories in the $2n$ dimensional
phase space, which is the vector space spanned by the
canonical variables. The Hamiltonian function is
assumed to be of the form
\beq
{\cal H} = \frac1{2m}\sum_j p_j^2 +
\frac12 \sum_{jk} K_{jk} q_jq_k,
\eeq
which corresponds to a collection of $n$ particles
of mass $m$ interacting harmonically.
The coefficients $K_{jk}$ are the entries of a
$n\times n$ matrix $K$ with positive eigenvalues.
Using the notations $q$ and $p$ for the column
matrices with elements $q_j$ and $p_j$, respectively,
we write 
\beq
{\cal H} = \frac1{2m} p^{\!\sf T} p
+ \frac12 q^{\!\sf T} K q.
\eeq

We perform a canonical transformation to write
the Hamiltonian function in a more symmetric form
in which it becomes invariant by the exchange
of the coordinate and momentum of a pair of
canonically conjugate variables. 
Taking into account that $K$ is Hermitian with
nonnegative eigenvalues we may define the matrix
$\Omega=K^{1/2}/\sqrt{m}$. It has nonnegative
eigenvalues and is Hermitian. The canonical 
transformation $q\to x$ and $p\to y$ is defined by
\beq
x = (m\Omega)^{1/2} q \qquad y = (m\Omega)^{-1/2} p,
\eeq
and the Hamiltonian function becomes
\[
{\cal H} = \frac1{2} y^{\!\sf T} \Omega y
+ \frac12 x^{\!\sf T} \Omega x,
\]
and we see that ${\cal H}$ is invariant by
the exchange of $x$ and $y$. In an explicit form,
\beq
{\cal H} = \frac1{2} \sum_{jk} \Omega_{jk}(x_jx_k+ y_jy_k),
\label{3a}
\eeq
and the Hamilton equations of motion becomes
\beq
\frac{dx_i}{dt} = \frac{\partial{\cal H}}{\partial y_i},
\qquad
\frac{dy_i}{dt} = -\frac{\partial{\cal H}}{\partial x_i}.
\eeq

Let us define the quantity
\beq
{\cal I} = \frac12\sum_j(x_j^2 + y_j^2).
\eeq
It follows
from the Hamilton equations that ${\cal I}$ is a constant
of the motion, a property that allows us to divide the
phase space in sectors, each one corresponding to a given
value of ${\cal I}$. We then {\it postulate} that the
only possible motions of the underlying system are those
corresponding to a defined value ${\cal I}$. This value
is denoted by $\mu$, 
\beq
\frac12\sum_j(x_j^2 + y_j^2) = \mu.
\label{5a}
\eeq
The constant $\mu$ has the physical dimension of
(energy$\times$time) and it will be seen that
$\mu$ is to be identified as the Planck constant. 
This postulate is the crucial step toward
quantization within the present approach. It enables
the appearance in the classical underlying system of
the characteristic quantum properties such as the
quantization of energy and the zero point of energy.

One can show that ${\cal I}$ is an adiabatic invariant
\cite{oliveira2022}, that is, a slow change of the
parameters of the Hamiltonian will change ${\cal H}$
but not ${\cal I}$. This result connects the adiabatic
invariance with quantization, an idea advanced by
Ehrenfest which lead to the old quantum mechanics.
The distinction between the approach of Ehrenfest
from ours lies in his use of the invariant
as connected to the motion of one quantum particle
as if the particle itself was a classical particle.

In fact ${\cal I}$ is not constant only in adiabatic 
changes but it is universally constant. Let us
suppose that ${\cal H}$ is time dependent,
that is, $\Omega_{jk}$ depends on time. 
In this case ${\cal H}$ is not a constant of motion
because  
\beq
\frac{d{\cal H}}{dt} = \frac12\sum_{jk}
\frac{d\Omega_{jk}}{dt} (x_j x_k + y_j y_k),
\eeq
but
\beq
\frac{d{\cal I}}{dt} = 0
\eeq 
This result allows to say that ${\cal I}$ remains
forever with the same value given at the initial
time, even if ${\cal H}$ is time dependent.

\subsection{Classical dynamics in complex variables}

The Hamilton equations of motion are written in terms
of a set of $2n$ complex variables $\phi_j$ and
$\phi_j^*$, $j=1,\ldots,n$, obtained by the
transformation
\beq
\phi_j =  \frac1{\sqrt{2\mu}}(x_j + i y_j),
\qquad
\phi_j^* = \frac1{\sqrt{2\mu}} (x_j - i y_j).
\eeq
Using this transformation, we see that the expression
(\ref{5a}) becomes
\beq
\sum_j \phi_j^*\phi_j = 1,
\label{6a}
\eeq
which expresses the postulate introduced above in
terms of the complex variables.

The peculiar transformation from real to complex
variables are canonical transformation leading
to the following Hamilton equation of motion
in complex variables
\beq
\frac{d\phi_j}{dt}
= \frac1{i\mu} \frac{\partial{\cal H}}{\partial\phi_j^*},
\qquad
\frac{d\phi_j^*}{dt}
= -\frac1{i\mu} \frac{\partial{\cal H}}{\partial\phi_j},
\label{58}
\eeq
where ${\cal H}$ is the real bilinear function 
\beq
{\cal H} = \sum_{jk} H_{jk}\phi_j^* \phi_k,
\label{56}
\eeq
obtained from (\ref{3a}), where
$H_{jk}=\mu (\Omega_{jk}+\Omega_{kj})/2$,
and are the entries of a $n\times n$ Hermitian matrix
$H$, that is, $H_{jk}^*=H_{kj}$, with positive eigenvalues.
We remark that the complex variables $\phi_j$ and 
$\phi_j^*$  are dimensionless, which justifies the
presence of the constant $\mu$ in the equations of motion.

The complex conjugate variables $\phi_j$ and
$\phi_j^*$ are considered to be independent variables
because their real and imaginary parts are proportional
to the coordinate $q_j$ and to the momentum $p_j$,
which are independent. From the peculiar transformation
above they also form a pair of canonically conjugate
variables as is manifest in the equation of motion
(\ref{58}).

From the Hamilton equations of motion (\ref{58}),
the time evolution of a state function ${\cal F}$ is
\beq
\frac{d{\cal F}}{dt} = \{{\cal F},{\cal H}\},
\eeq
where the term in the right-hand side is the
Poisson brackets defined by
\beq
\{{\cal F},{\cal H}\} = \sum_j(
\frac{\partial{\cal F}}{\partial\phi_j}
\frac{\partial{\cal H}}{\partial\phi_j^*}
- \frac{\partial{\cal H}}{\partial\phi_j}
\frac{\partial{\cal F}}{\partial\phi_j^*}),
\eeq
If we replace $\cal F$ by the norm of $\phi$,
defined by
\beq
{\cal N} = \sum_j \phi_j^* \phi_j,
\label{57}
\eeq
we see that 
\beq
\{{\cal N},{\cal H}\} = 0,
\label{7}
\eeq
which means that the norm is preserved along a
trajectory in the complex vector space.

The value of ${\cal N}$ cannot be arbitrary. In
accordance with the postulate introduced above and
expressed by equation (\ref{6a}), it should be equal
to the unity. Using this result and writing the axes
corresponding to a pair of canonical conjugate variables 
of the phase space as related to a complex variable
$\phi_j$, then the phase space becomes equivalent
to a Hilbert space, which is a vector space with
normalized vectors.

The form (\ref{56}) of the Hamiltonian allows us to
write the Hamilton equations in the form
\beq
\frac{d\phi_j}{dt}
= \frac1{i\mu} \sum_k H_{jk}\phi_k.
\label{58a}
\eeq
Equivalently, we may
write equation (\ref{58a}) in the vector form
\beq
i\mu\frac{d\phi}{dt} =  H \phi.
\label{58b}
\eeq
If we set $\mu=\hbar$ and identify $\phi$ as the
quantum state vector, we see that it is identical to
the quantum equation that gives the time evolution of
the state vector $\phi$, or the Schrödinger equation.

\subsection{Norm preserving noise}

We now assume that $\phi$ follows a stochastic equation
of motion which is the Hamilton equations supplemented
by a noise term,
\beq
\frac{d\phi_j}{dt}
= \frac1{i\mu} \sum_{jk} H_{jk}\phi_k,
 + \zeta_j,
\label{3}
\eeq 
where $\zeta_j$ is a complex white noise, that is,
its real and imaginary parts are white noises, and also
depends on $\phi$ and $\phi^*$. The noise
will be set up in such a way that the norm ${\cal N}$
is preserved along the stochastic trajectory in 
the complex vector space.

As it stands equation (\ref{3}) has no precise meaning.
To give it a precise meaning, we discretize the time
in intervals $\tau$ and write the stochastic equation
of motion in a discretized form. Let $\Delta \phi_j$
be the increment in $\phi_j$ when the time increases
from $t$ to $t+\tau$. Then the discretized version
of equation (\ref{3}) is assumed to be
\beq
\Delta \phi_j =  \frac\tau{i\mu}\sum_k H_{jk}\phi_k
+ i \sqrt{\tau} \sum_k G_{jk}\phi_k
- \frac{\tau}2 \sum_k K_{jk}\phi_k,
\label{19}
\eeq
where $G_{jk}$ are independent random variables, which
we consider to be the elements of a $n\times n$ matrix
$G$, with zero mean, $\la G_{jk}\ra=0$, and covariances
\beq
\la G_{jk}^* G_{\ell m}\ra = 2\gamma_{jk,\ell m}.
\label{25}
\eeq
from which follows the property
$\gamma_{jk,\ell m}^*=\gamma_{\ell m,jk}$.
The quantities $K_{jk}$ are related to $G_{jk}$ and
will be found by imposing the conservation of the norm.
The elements $G_{jk}$ and $K_{jk}$ are understood
as the entries of two $n\times n$ matrices
$G$ and $K$, respectively.

It is convenient to introduce a $n^2\times n^2$
matrix $\Gamma$ with elements $\Gamma_{rs}$ which are
related to $\gamma_{jk,\ell,m}$. The index $r$ has
a one-to-one correspondence to $(j k)$ and $s$ has
a one-to-one correspondence to $(\ell m)$. Using this
convention, set $\Gamma_{rs}=\gamma_{r,s}$, or
\beq
\Gamma_{rs} = \la G_r^* G_s\ra.
\label{10}
\eeq
From this relation it follows that $\Gamma$ is a
Hermitian matrix, $\Gamma_{rs}^*=\Gamma_{sr}$. Being
Hermitian, $\Gamma$ can be diagonalized by a unitary
transformation. Denoting by $\Upsilon$ the
$n^2\times n^2$ matrix that diagonalizes $\Gamma$
then $\Upsilon^\dagger\Gamma\Upsilon$ is diagonal
and its elements $\lambda_r$ are the eigenvalues
of $\Gamma$. They are given by
\beq
\lambda_r = \sum_{s's} \Upsilon^\dagger_{rs'}
\Gamma_{s's}\Upsilon_{sr}.
\eeq
Using (\ref{10}),
\beq
\lambda_r = \sum_{s's} \Upsilon^\dagger_{rs'}
\la G^*_{s'} G_s\ra\Upsilon_{sr}
= \la |\sum_s G_s\Upsilon_{sr}|^2\ra,
\eeq
and we see that $\lambda_r\geq 0$, that is the
eigenvalues of the Hermitian matrix $\Gamma$ are
nonnegative, or in other words, $\Gamma$ is a
positive semi-definite matrix.

Let us determine the increment in the norm
${\cal N}$ due to a change $\Delta\phi_j$
in the dynamic variables. It is given by
\beq
\Delta {\cal N} = \sum_j(\Delta\phi_j^* \phi_j
+ \phi_j^* \Delta\phi_j + \Delta\phi_j^* \Delta\phi_j).
\eeq
Replacing $\Delta \phi_j$ in this equation,
we find up to terms of order $\tau$
\[
\Delta {\cal N} = 
i \sqrt{\tau} \sum_{jk} (G_{jk}-G_{kj}^*)\phi_j^*\phi_k
\]
\beq
- \frac{\tau}2\sum_{jk} (K_{jk}+K_{kj}^*)\phi_j^* \phi_k
+ \tau\sum_{jk} (G^\dagger G)_{jk} \phi_j^*\phi_k.
\label{34a}
\eeq
The terms containing $H_{jk}$ vanish identically due
to the Hermitian property $H_{jk}^*=H_{kj}$. Choosing
\beq
K=G^\dagger G,
\eeq
which is the sought relation between $K$ and $G$,
and which we assume from now on, then the last two
summations on the right-hand side of (\ref{34a})
vanish and 
\beq
\Delta{\cal N} = i\sqrt{\tau}\sum_{jk}
\phi_j^* (G - G^\dagger)_{jk}\phi_k.
\eeq
If $G^\dagger=G$ then the increment vanishes and the
norm $\cal N$ is strictly constant along the stochastic
trajectory. If this condition is not imposed, the
increment in the norm will still vanish but in the
average, that is, $\la\Delta{\cal N}\ra=0$.
The stochastic equation of motion (\ref{19})
defines a Markovian stochastic dynamics which
determines stochastic trajectories of $\phi$
in the vector space.

\section{Fundamental equation}

\subsection{Probability density distribution}

As the trajectories in the vector space are stochastic,
we may ask for the probability of the occurrence of
each trajectory. In the following we determine the
equation that gives the time evolution of the
probability density distribution
${\cal P}(\phi,\phi^*,t)$ of $\phi$ and $\phi^*$
at time $t$.
We start by considering an arbitrary state function
${\cal F}$ of $\phi$ and $\phi^*$ of the bilinear 
type
\beq
{\cal F} = \sum_{jk} F_{jk}\phi_j^* \phi_k,
\label{1}
\eeq
where $F_{jk}$ are understood as the complex entries
of a $n\times n$ matrix $F$. The increment
$\Delta{\cal F}$ of such a function is given by
\[
\Delta{\cal F}
= \sum_j \frac{\partial{\cal F}}{\partial\phi_j}
\Delta\phi_j + \sum_j \frac{\partial{\cal F}}
{\partial\phi_j^*}\Delta\phi_j^*
\]
\beq
+ \sum_{jk} \frac{\partial^2{\cal F}}
{\partial\phi_j\partial\phi_k^*}
\Delta\phi_j\Delta\phi_k^*.
\label{37}
\eeq
Replacing $\Delta \phi_j$ in this equation we
find up to terms of order $\tau$
\[
\Delta{\cal F} = \frac{\tau}{i\mu}\{{\cal F},{\cal H}\}
+ i\sqrt{\tau} \sum_{jk}
(\frac{\partial{\cal F}}{\partial\phi_j} G_{jk}\phi_k
-\frac{\partial{\cal F}}{\partial\phi_k^*}
G_{kj}^*\phi_j^*)
\]
\[
- \frac{\tau}2 \sum_{jk}
(\frac{\partial{\cal F}}{\partial\phi_j} K_{jk}\phi_k
+ \frac{\partial{\cal F}}{\partial\phi_k^*} K_{kj}^*
\phi_j^*)
\]
\beq
+ \tau \sum_{jk\ell m}
\frac{\partial^2{\cal F}}{\partial\phi_j\partial\phi_k^*}
G_{j\ell} G_{km}^*\phi_\ell \phi_m^*.
\label{35}
\eeq

Taking the average of both sides of equation
(\ref{35}), the term proportional to $\sqrt{\tau}$ 
vanishes. After that, we divide what is left by
$\tau$ to reach the result
\[
\frac{d}{dt}\la{\cal F}\ra
= \frac1{i\mu} \la\{{\cal F},{\cal H}\}\ra
- \sum_{jk\ell}\gamma_{\ell j,\ell k} 
(\la\phi_k\frac{\partial{\cal F}}{\partial\phi_j}\ra
+ \la\phi_j^*\frac{\partial{\cal F}}{\partial\phi_k^*}\ra)
\]
\beq
+ 2\sum_{jk\ell m}\gamma_{km,j\ell}\la\phi_\ell \phi_m^*
\frac{\partial^2{\cal F}}
{\partial\phi_j\partial\phi_k^*}\ra.
\label{36}
\eeq
where here the average are taken over the probability
density distribution ${\cal P}$, that is,
the average $\la{\cal F}\ra$ of ${\cal F}$ is
\beq
\la {\cal F}\ra = \int {\cal F}{\cal P}d\phi d\phi^*.
\label{22}
\eeq

Taking into account that ${\cal F}$ is an arbitrary
function, we reach the equation for the time 
evolution of the probability distribution
${\cal P}$, which is
\[
\frac{\partial{\cal P}}{\partial t}
= \frac1{i\mu} \{{\cal H},{\cal P}\}
+ \sum_{jk\ell} \gamma_{\ell j,\ell k}
(\frac{\partial \phi_k{\cal P}}{\partial\phi_j}
+ \frac{\partial\phi_j^*{\cal P}}{\partial\phi_k^*})
\]
\beq
+ 2\sum_{jk\ell m} \gamma_{km,j\ell} 
\frac{\partial^2 \phi_\ell \phi_m^*{\cal P}}
{\partial\phi_j\partial\phi_k^*}.
\label{39}
\eeq
To reach this equation we bear in mind that the averages 
in (\ref{36}) are integrals in the complex vector space
of the type (\ref{22}). The expressions in (\ref{39}) are
found performing appropriate integrals by parts and
considering that ${\cal P}$ vanishes rapidly at the
boundaries of integration. The equation (\ref{39})
is recognized as a Fokker-Planck-Kramers equation
\cite{vankampen1981,risken1989,gardiner2009,tome2015},
in several complex variables.

\subsection{Master equation}

The fundamental equation (\ref{39}) was derived above 
considering a discrete time stochastic equation of
motion and then taking the continuous time limit.
Here we consider another derivation of the fundamental
equation by considering a continuous time 
equation of motion but discretized variables $\phi$
and $\phi^*$. 
To simplify the notation we
write $\varphi$ in the place of $(\phi,\phi^*)$
and consider the following Kolmogorov equation,
or master equation,
\beq
\frac{d}{dt}{\cal P}(\varphi) = \sum_{\varphi'}
\{W(\varphi,\varphi'){\cal P}(\varphi')
- W(\varphi',\varphi){\cal P}(\varphi)\},
\eeq
where $W(\varphi',\varphi)$ are the entries of a
stochastic matrix $W$. A stochastic matrix holds two
properties: 1) the off diagonal
entries $W(\varphi',\varphi)$ are nonnegative and
represent the probability rate of the transition
$\varphi\to \varphi'$, and 2) it holds the property
\beq
\sum_\varphi W(\varphi,\varphi') = 0.
\label{33}
\eeq
From these two properties, it follows 
from the Perron-Frobenius theorem that
${\cal P}(\varphi,t)\geq0$ if at the initial time
it holds this property, and it is normalized
at all times,
\beq
\sum_\varphi{\cal P}(\varphi,t) = 1.
\eeq 

It is more convenient to construct the backward
Kolmogorov equation, or adjoint master equation,
\beq
\frac{d}{dt}{\cal Q}(\varphi)
= \sum_{\varphi'} W(\varphi,\varphi')
\{{\cal Q}(\varphi') - {\cal Q}(\varphi)\}.
\label{40}
\eeq

We consider three types of transitions.
The first is defined by 
\beq
\phi_j \to \phi_j'=\phi_j+ \frac{\varepsilon}{i\mu}
H_{j\ell}\phi_\ell,
\eeq
\beq
\phi_k^* \to \phi_k^{\prime*}=\phi_k^* -
\frac{\varepsilon}{i\mu} H_{mk}\phi_m^*,
\eeq
and occurs with rate one. The contribution of
this transition to the right-hand side of equation
(\ref{40}) is given by the expression
\beq
\frac1{i\mu}\sum_{j\ell}\{H_{j\ell}\phi_\ell
\frac{\partial {\cal P}}{\partial\phi_j} 
- H_{\ell j}\phi_\ell^* 
\frac{\partial{\cal P}}{\partial\phi_j^*}\}.
\eeq

The second type is defined by
\beq
\phi_j \to \phi_j'= \phi_j \pm i\sqrt\varepsilon
g_{j\ell}\phi_\ell,
\eeq
\beq
\phi_k^* \to \phi_k^{\prime*} = \phi_k^*
\mp i\sqrt\varepsilon g_{km}^*\phi_m^*,
\eeq
and occurs with transition rate $\alpha_{\ell m,jk}\geq0$,
which holds the property
\beq
\alpha_{\ell m,jk}=\alpha_{m\ell,kj}.
\label{51}
\eeq
The contribution of this transition to the right-hand
side of equation (\ref{40}) is given by the expression
\beq
2\sum_{jk\ell m}\alpha_{\ell m,jk}
\frac{\partial^2{\cal Q}}{\partial\phi_j\partial\phi_k^*}
g_{j\ell}\phi_\ell g_{km}^*\phi_m^*,
\eeq
which is real due to the property (\ref{51}).

The third type is defined by
\beq
\phi_j \to \phi_j'= \phi_j
- \frac12\varepsilon g^*_{\ell j} g_{\ell k}\phi_k,
\eeq
\beq
\phi_k^* \to \phi_k^{\prime*} = \phi_k^*
- \frac12\varepsilon g^*_{\ell j} g_{\ell k}\phi_j^*,
\eeq
occurring with rate $\alpha_{kj,\ell\ell}\geq0$.
The contribution of this transition to the right-hand
side of equation (\ref{40}) is given by the expression
\[
-\sum_\ell \alpha_{kj,\ell\ell} g^*_{\ell j} g_{\ell k}\{
\frac{\partial {\cal Q}}{\partial\phi_j}\phi_k
+ \frac{\partial{\cal Q}}{\partial\phi_k^*}\phi_j^*\},
\]
which is real due to the property (\ref{51}).

Replacing the expressions obtained above
in the right-hand side of equation (\ref{40}),
it becomes identical to the adjoint equation
(\ref{36}), if we set  
\beq
\gamma_{km,j\ell} = \alpha_{\ell m,jk} g_{j\ell}g_{km}^*. 
\label{49}
\eeq

Taking into account that $\gamma_{km,j\ell}$
is the covariance $\la G_{km}^*G_{j\ell}\ra$,
then the random variables $G_{jk}$ must obey
the relation 
\beq
\la G_{km}^*G_{j\ell}\ra 
= \alpha_{\ell m,jk} g_{j\ell}g_{km}^* .
\label{49a}
\eeq
This is accomplished if the absolute value of
$G_{jk}$ is a random variable but not its
phase. Indeed, if we write $G_{jk}=R_{jk}e^{i a_{jk}}$
then
\beq
\la G_{km}^*G_{j\ell}\ra =  \la R_{km} R_{j\ell}\ra
e^{- i a_{km}} e^{i a_{j\ell}},
\eeq
which is of the form (\ref{49a}).

\subsection{Central equation}

Equation (\ref{39}) that governs the time evolution of the
probability density distribution $\mathscr{P}(\phi,\phi^*)$
is the fundamental equation of the present stochastic
approach. From this equation we derive the equation for 
the time evolution of any bilinear state function
such as th covariances $\rho_{jk}=\la\phi_j\phi_k^*\ra$
of the stochastic variables $\phi_j$. The equation that
give the time evolution of $\rho_{jk}$ is obtained from
(\ref{39}) but we may as well use the equation
(\ref{36}) by replacing ${\cal F}$ by
$\phi_j\phi_k^*$. The result is
\[
\frac{d}{dt}\rho_{jk} =\frac1{i\mu} \sum_\ell(
H_{j\ell} \rho_{\ell k} - \rho_{j\ell} H_{\ell k})
\]
\beq
- \sum_{\ell m}
(\gamma_{mj,m\ell} \rho_{\ell k}
+ \rho_{j\ell}\gamma_{m\ell,mk})
+ 2\sum_{\ell m}\gamma_{km,j\ell} \rho_{\ell m}.
\label{55}
\eeq
Once this equation is solved, the average
$\la{\cal F}\ra$ of a bilinear function  ${\cal F}$
is obtained by
\beq
\la{\cal F}\ra = \sum_{jk} F_{jk}\rho_{kj}.
\label{61}
\eeq

Equation (\ref{55}) is the central equation of the
present approach and is the most general form of an
equation for the covariance $\rho_{jk}$ that can be
derived from a noise that is linear in $\phi_j$ and
which conserves in the average the norm of $\phi_j$.
The coefficients $\gamma_{\ell j,km}$ are not
arbitrary. As mentioned above,
$\gamma_{\ell j,km}^*=\gamma_{km,\ell j}$
and the eigenvalues of the $n^2\times n^2$ matrix
with elements $\gamma_{\ell j,km}$ are nonnegative.

The central equation (\ref{55}) can be written in 
matrix form as follows. We first define the covariance
matrix $\rho$ as the $n\times n$ matrix with elements
$\rho_{jk}$. Then we introduce $n\times n$ matrices
$A^{jk}$ whose entries are all zero except the entry
at row $j$ and column $k$ which equals 1. Notice that
$A^{jk}$ is not an entry of a matrix but denotes one
of a collection of $n^2$ matrices. Their entries are
denoted by $A^{jk}_{\ell m}$ and are given by
\beq
A^{jk}_{\ell m} = \delta_{j\ell}\delta_{km}.
\eeq
The matrices $A^{jk}$ form a complete basis for
the expansion of any $n\times n$ matrix. For
instance, in terms of this set the matrix
$\rho$ has the expansion
\beq
\rho = \sum_{jk} \rho_{jk}A^{jk}.
\eeq

In terms of the basis matrices, the central
equation (\ref{55}) becomes
\[
\frac{d\rho}{dt} =\frac1{i\mu} [H,\rho]
\]
\beq
+ \sum_{jk\ell m}\!\gamma_{jk,\ell m}
(2A^{\ell m} \rho A^{jk\dagger} 
- A^{jk\dagger} A^{\ell m} \rho
- \rho A^{jk\dagger} A^{\ell m}),
\label{24a}
\eeq
where $[H,\rho]=H\rho-\rho H$ stands for the
commutation between the matrices $H$ and $\rho$.

Using (\ref{61}), the average of $\la{\cal F}\ra$
of a bilinear state function ${\cal F}$ is
determined from $\rho$ by
\beq 
\la{\cal F}\ra = {\rm Tr}F\rho.
\eeq

The matrix $\rho$ holds the following properties.
It is a Hermitian matrix with unit trace,
\beq
{\rm Tr}\rho = 1,
\eeq
and is a semi-positive definite matrix, which means that
its eigenvalues are non-negative. These properties, which
allow us to call $\rho$ a density matrix, follow from the
definition of $\rho$ as a covariance matrix, that is,
\beq
\rho_{jk}= \int \phi_j\phi_k^* \mathscr{P}d\phi d\phi^*,
\label{50}
\eeq
and from the properties of the distribution density 
which are $\mathscr{P}\geq0$ and normalization,
\beq
\int \mathscr{P}d\phi d\phi^* = 1.
\eeq
The Hermitian property $\rho_{jk}^*=\rho_{kj}$ follows
from (\ref{50}). As $\rho$ is Hermitian its eigenvalues
are real. To show that the eigenvalues are nonnegative
it suffices to consider a transformation that diagonalizes
$\rho$ and use this transformation to  change the
variables $\phi_j$ to new variables $\phi_j'$. Taking
into account that $\rho_{jk}=\la \phi_j\phi_k^*\ra$ then 
$\la\phi_j'\phi_k^{\prime *}\ra$ will be diagonal and
coincides with the eigenvalues $p_j$ of $\rho$.
Therefore, $p_j=\la\phi_j'\phi_j^{\prime *}\ra\geq 0$.

The fundamental equation was construct in such a
way that the norm was conserved in the average,
which means that 
\beq
\la {\cal N}\ra = \sum_j \rho_{jj}
\label{23}
\eeq
is constant. Choosing this constant to be equal 
to unity, then ${\rm Tr}\rho=1$. Alternatively,
it follows from equation (\ref{24a}) that
$d\,{\rm Tr}\rho/dt=0$. The semi-definite property
of $\rho$ is also preserved at all times because
at any time $\rho_{jk}$ keeps being a covariance.

The above properties are valid as long as $\mathscr{P}$
conserves the properties of a probability density
distribution stated above. But this is indeed the
case as we have demonstrated above.

The equation (\ref{24a}) is formally identical to the 
quantum master equation \cite{breuer2002,manzano2020}
introduced by Lindblad \cite{lindblad1976} and by
Gorini, Kossakowski, and Sudarshan \cite{gorini1976},
describing the time evolution of a density matrix of
a quantum open system. The distinguish feature between
them is that equation (\ref{24a}) was obtained by 
considering that $\rho$ is a covariance matrix. 
Of course, as we have shown above, it turned out
to be a density matrix.

It is worth considering the time evolution of the
average $\chi_j=\la\phi_j\ra$. From (\ref{36})
we find 
\beq
\frac{d\chi_j}{dt} =\frac1{i\mu} \sum_k H_{jk}\chi_k
- \frac12 \sum_k \gamma_{jk} \chi_k,
\label{43}
\eeq
where $\gamma_{jk}=\la K_{jk}\ra$.

\section{Quantum mechanics}

The noise defined by the last two terms of (\ref{19})
changes the variable $\phi_j$ by changing both
the phase $\theta_j$ and the absolute value $r_j$ of
$\phi_j=r_je^{i\theta_j}$. However, if the noise is
of the type given by the equation
\beq
\Delta \phi_j =
(i\sqrt{\tau} \xi - \frac{\tau}2 \gamma)\phi_j,
\label{26}
\eeq
where $\xi$ is a random variable with zero mean and
variance $\gamma$, then the noise still changes the
phase $\theta_j$ but not the absolute value of
$\phi_j$. To show this result it suffices to
write this equation in the equivalent form
\beq
\phi_j'
= e^{i\sqrt{\tau} \gamma \xi} \phi_j,
\eeq
where $\phi_j' = \phi_j + \Delta \phi_j$. This result
shows that $\phi_j^*\phi_j=r_j^2$ is invariant and
the norm ${\cal N}$ is conserved in the strict sense.

The noise defined by the expression (\ref{26})
corresponds to set
$2\gamma_{j\ell,k m} = \gamma\delta_{j\ell}\delta_{km}$.
The equation (\ref{36}) reduces to  
\[
\frac{d}{dt}\la{\cal F}\ra
= \frac1{i\mu} \la\{{\cal F},{\cal H}\}\ra
\]
\beq
- \frac12\gamma \sum_j
\la \phi_j\frac{\partial{\cal F}}{\partial\phi_j}
+ \phi_j^*\frac{\partial{\cal F}}{\partial\phi_j^*}\ra
+ \gamma \sum_{jk} \la\phi_j \phi_k^*
\frac{\partial^2{\cal F}}
{\partial\phi_j\partial\phi_k^*}\ra,
\eeq
and the fundamental equation (\ref{39}) becomes
\[
\frac{d\mathscr{P}}{dt}
= \frac1{i\mu} \{{\cal H},\mathscr{P}\}
\]
\beq
+ \frac{\gamma}2\sum_j
(\frac{\partial \phi_j\mathscr{P}}{\partial\phi_j}
+ \frac{\partial\phi_j^*\mathscr{P}}{\partial\phi_j^*})
+ \gamma\sum_{jk}\frac{\partial^2 \phi_j
\phi_k^*\mathscr{P}} {\partial\phi_j\partial\phi_k^*},
\eeq
which reduces to the simpler form
\beq
\frac{\partial\mathscr{P}}{\partial t} 
= \frac1{i\mu}\{\mathscr{P},{\cal H}\}
+ \gamma^2 \sum_{jk}
\frac{\partial^2 \mathscr{P}}
{\partial\theta_j\partial\theta_k}.
\label{39a}
\eeq
In this case the equation (\ref{55}) becomes
\beq
\frac{d}{dt}\rho_{jk} =\frac1{i\mu} \sum_\ell(
H_{j\ell} \rho_{\ell k} - \rho_{j\ell} H_{\ell k}),
\label{21}
\eeq
which can be written in the matrix form as
\beq
\frac{d\rho}{dt} = \frac1{i\mu} [H,\rho],
\label{21a}
\eeq
which is the quantum Liouville equation, if we set
$\mu=\hbar$. 
The equation (\ref{43}) for $\chi_j$ becomes
\beq
\frac{d\chi_j}{dt} =\frac1{i\mu} \sum_k H_{jk}\chi_k
- \frac12 \gamma \chi_j.
\label{43a}
\eeq
We remark that the equation (\ref{21}) does not have
the terms corresponding to the noise. However, this
does not mean that the effect of the noise is not
present. The variable $\phi_j$ is still a stochastic
variable which is reflectedin the last term of
equation (\ref{43a}).

The equation (\ref{21}) has a special type of solution
which is $\rho_{jk}=\psi_j\psi_k^*$. Replacing this
form of $\rho$ in the equation (\ref{21}), we see that
it is indeed a solution as long as  $\psi_j$
fulfills the equation 
\beq
\frac{d\psi_j}{dt} = \frac1{i\mu} \sum_k H_{jk} \psi_k,
\label{43b}
\eeq
or in matrix form
\beq
\frac{d\psi}{dt} = \frac1{i\mu} H \psi,
\label{43c}
\eeq
where $\psi$ is a column matrix with elements $\psi_j$,
and
\beq
\rho = \psi \psi^\dagger,
\eeq
were $\psi^\dagger$ is a row matrix with elements
$\psi_j^*$. A solution of this type is called pure
state and equation (\ref{43c}) is identified as
the Schrödinger equation, if we set $\mu=\hbar$.
We remark that $\psi_j$ should not be confused with
$\phi_j$ nor with $\chi_j=\la\phi_j\ra$.  This last
quantity obeys the equation (\ref{43a}) which
differs from (\ref{43b}) by the presence of
the term $-\gamma\chi_j/2$ which makes
$\chi_j$ to vanish in the long term.  

The relation between $\phi_j$ and $\psi_j$ is
\beq
\psi_j\psi_k^*=\la\phi_j\phi_k^*\ra=\rho_{jk},
\eeq
and in particular 
\beq
|\psi_j|^2=\la|\phi_j|^2\ra=\rho_{jj}.
\eeq
Due to the invariance of the norm given by (\ref{23}) 
it follows the usual normalization of $\psi_j$,
\beq
\sum_j |\psi_j|^2 =1.
\label{62}
\eeq 
The average $\la{\cal F}\ra$ given by
equation (\ref{61}) now reads
\beq
\la{\cal F}\ra = \sum_{jk} \psi_k^* F_{jk}\psi_j,
\label{61a}
\eeq
which is the usual expression for the quantum average.

\section{Quantum Thermodynamics}

\subsection{First law of thermodynamics}

We start by writing the central equation (\ref{24a})
in a more convenient form as follows 
\beq
\frac{d\rho}{dt} =\frac1{i\mu} [H,\rho] + L,
\qquad L = \sum_r L^r
\label{69}
\eeq
where
\[
L^r = \frac12\sum_s
\{\gamma_{r,s} [A^s\rho,A^{r\dagger}]
+ \gamma_{r,s}'[A^s \rho,A^r]
\]
\beq
+ \gamma_{r,s}^*[A^r,\rho  A^{s\dagger}]
+ \gamma_{r,s}^{\prime *} [A^{r\dagger},\rho
A^{s\dagger}]\},
\eeq
and we are using a convention concerning the index
notation defined by the replacements $jk\to r$ and
$\ell m\to s$, and
$\gamma_{jk,\ell m}'=\gamma_{kj,\ell m}$,
and we remark that $L^{r\dagger}=L^r$.
Defining the auxiliary matrices
\beq
B^r = \sum_s \gamma_{rs} A^s, \qquad
C^r = \sum_s \gamma_{r,s}' A^s,
\label{59}
\eeq
then $L^r$ acquires the simpler form
\beq
L^r = \frac12\{[A^r,\rho B^{r\dagger} - C^r\rho]
- [A^{r\dagger}, B^r\rho -\rho C^{r\dagger}]\}.
\label{24c}
\eeq
Notice that $B^{jk}_{\ell m}=\gamma_{jk,\ell m}$
and $C^{jk}_{\ell m}=\gamma_{kj,\ell m}$.

From the central equation (\ref{69}) we determine
the equation that gives the time evolution of the
average $\la{\cal F}\ra={\rm Tr}(F\rho)$ of a
bilinear state function. It is given by
\beq
\frac{d\la{\cal F}\ra}{dt}
= \frac1{i\mu} {\rm Tr} F [H,\rho]
+ \sum_r{\rm Tr} F L^r.
\label{38}
\eeq

In accordance with
thermodynamics the change in the energy ${\cal E}$
of an open system equals the heat introduced into
the system minus the work done by the system
on the environment, which is the law of conservation
of energy, or the first law of thermodynamics.
The work is assumed to be the increment in
a potential ${\cal V}$ due to the external forces. 
The conservation of energy is then written as 
\beq
\frac{d\la{\cal E}\ra}{dt}
= \Phi - \frac{d\la{\cal V}\ra}{dt},
\eeq
where $\Phi$ is the flux of heat into the system.
We assume that the Hamiltonian function ${\cal H}$
is the sum of the energy function ${\cal E}$ and
the potential ${\cal V}$ of the external forces,
${\cal H}={\cal E}+{\cal V}$.
Bearing this in mind, we write
\beq
\frac{dU}{dt} = \Phi,
\label{45}
\eeq
where $U=\la{\cal H}\ra$.

To determine the expression of $\Phi$, we calculate
the left-hand side of equation (\ref{45}) using
equation (\ref{38}). The result for the flux is
\beq
\Phi = \sum_r \Phi_r,
\label{28a}
\eeq
where 
\beq
\Phi_r = {\rm Tr} H L^r.
\label{28}
\eeq
Equation (\ref{45}), with $\Phi$ representing
the total flux of heat into the system,
expresses the conservation of energy, or the first
law of thermodynamics. 

\subsection{Second law of thermodynamics}

If we wish to describe an open system regarded
as a thermodynamic system in the sense that it
obeys the laws of thermodynamics we need to
introduce two quantities which are the entropy
$S$ of the system and the flux of entropy $\Psi$. 

Let $f(x)\leq0$ be a decreasing and convex function
of $x$ defined on the interval $0\leq x\leq 1$, and
such that $f(1)=0$. We also require that 
$xf(x)\to0$ when $x\to0$. An example of  $f(x)$
holding these properties is the function $-\ln x$.
The entropy $S$ of the system is then defined by
\beq
S = \kappa \sum_j p_j f(p_j),
\label{47}
\eeq
where $\kappa$ is some positive constant and
$p_j\geq0$ are the eigenvalues of the matrix $\rho$,
which are subject to the condition
\beq
\sum_j p_j = 1,
\eeq
following from ${\rm Tr}\rho=1$. 

Considering that $0\leq p_j\leq 1$ it follows that
$S\geq0$. The minimum value of the entropy is $S=0$
which occurs when one of the quantities $p_j$ equals
one, which is the case of pure states. In terms of
the matrix $\rho$, the entropy can be written as
\beq
S = \kappa \,{\rm Tr}\rho  f(\rho).
\eeq
Deriving $S$ with respect to time and using the
central equation (\ref{69}), we find
\beq
\frac{dS}{dt} = \kappa \sum_r {\rm Tr} L^r f(\rho).
\label{27}
\eeq
By calling $S$ the entropy we wish that it describes
the thermodynamic entropy. A distinguishing property
of thermodynamic entropy is that it is not a conserved
quantity. For instance, a system that has its energy
increased only because of heat flow must have its
entropy increased. Therefore, the right-hand side of
(\ref{27}) is not in general the entropy flux.

Concerning the entropy flux, we assume with Clausius
that the entropy flux is the ratio between the heat
flux and the temperature of the environment. More
precisely, we assume that the total entropy flux is
a sum of entropy fluxes coming from distinct parts
of the environment which are at distinct temperatures.
The entropy flux $\Psi_r$ coming from the $r$-th part
of the environment at temperature $T_r$ is
\beq
\Psi_r = \frac{\Phi_r}{T_r},
\label{34}
\eeq
where $\Phi_r$ is the heat flux coming from the $r$-th
part which we assume to be given by the expression
(\ref{28}). However, we cannot adopt the Clausius
relation in the strict sense because no temperature
has yet been defined. But we can still assume the
Clausius relation, with the understanding that
$T_r>0$ are parameters of the present approach.

Replacing the expression for the heat flux given
by (\ref{28}) in (\ref{34}), we find the desired
expression for the flux of entropy coming from
the $r$-th part of the environment,
\beq
\Psi_r = \frac1{T_r}{\rm Tr} H L^r.
\label{20}
\eeq
The total flux of entropy is
\beq
\Psi = \sum_r \Psi_r.
\label{20a}
\eeq

The right-hand side of (\ref{27}) must be the sum
of the entropy flux $\Psi$ from the environment and
another term corresponding to the creation of
entropy which is the rate of entropy production
$\Pi$ within the system, defined by
\beq
\frac{dS}{dt} = \Pi + \Psi,
\label{42}
\eeq
From this relation we determine $\Pi$ using
the expressions of the flux of entropy $\Psi$
and $dS/dt$. Before that, we define $\rho^r$ 
through the relation
\beq
\beta_r H = f(\rho^r) + c_r,
\label{63}
\eeq
where $c_r$ is such that ${\rm Tr}\rho_r=1$,
and $\beta_r=1/\kappa T_r$. 
From this relation and equation (\ref{20})
we may write $\Psi_r$ as
\beq
\Psi_r = \kappa {\rm Tr} L^r f(\rho^r),
\label{18}
\eeq
where the constant $C_r$ has disappeared because
${\rm Tr}L^r=0$. Subtracting (\ref{18}) from
(\ref{27}), we find 
\beq
\Pi = \kappa \sum_r {\rm Tr} L^r \{f(\rho) - f(\rho^r)\}.
\label{74}
\eeq

According to Clausius the increase in the entropy
$S$ of a system is larger than or equal to the
entropy flux $\Psi$ into the system,
\beq
\frac{dS}{dt} \geq \Psi,
\label{48}
\eeq 
which constitutes the second law of thermodynamics.
Taking into account the equality (\ref{42}), the
Clausius expression (\ref{48}) of the second law 
can be written in the equivalent form
\beq
\Pi \geq 0.
\eeq
In the following, we will demonstrate this inequality
proving thus the Clausius expression for the second
law of thermodynamics.

\subsection{Positivity of the entropy production}

To show that $\Pi\geq0$, we demonstrate that each
term of the summation in (\ref{74}) is nonnegative,
that is, we demonstrate that $\Pi_r\geq0$,
\beq
\Pi_r = \kappa{\rm Tr} L^r(\rho) \{f(\rho)- f(\rho^r)\},
\eeq
and $L^r$ is given by (\ref{24c}).

Our procedure is to show (a) that $\Pi_r$ is a convex
function of the eigenvalues $p_\ell$ of $\rho$, and 
(b) that $\Pi_r$ is bounded from below at $\rho=\rho^r$,
in which case $\Pi_r=0$.
If the first proposition is demonstrated, then
the second proposition can be met if $\rho=\rho^r$
is a double zero of $\Pi_r$ which amounts to say that
$L^r$ vanishes when $\rho=\rho^r$. This condition
leads consistently to the vanishing of $\Psi_r$
as well. The vanishing of $L^r$ is fulfilled by
demanding that
\beq
B^r\rho^r = \rho^r C^{r\dagger},
\label{54}
\eeq
which we assume from now on. As $B^r$ and $C^r$ are
defined in terms of $\gamma_{r,s}$ and $\rho^r$
depends on $T_r$, this condition establishes a
relation between the correlations $\gamma_{r,s}$
and the parameter $T_r$.

To show that $\Pi_r$ is convex it suffices to
show that
\beq
R_r = {\rm Tr} L^r(\rho) f(\rho)
\eeq
is convex because $L^r(\rho)$ is linear in $\rho$.
We provide the demonstration of convexity of
$R_r$ for the case in which $L(\rho)$ has a
diagonal form, which is equivalent to express
$L^r(\rho)$ as 
\beq
L^r =  \frac{\lambda_r}2\{[A^r,[\rho, A^{r\dagger}]]
+ [A^{r\dagger}, [\rho, A^r]]\},
\eeq
in which case
\beq
R_r = \lambda_r \Re {\rm Tr}
\{[f,A^{r\dagger}\rho]A^r  +[f,A^r\rho] A^{r\dagger} \}.
\eeq

Next, we consider the following term of $R_r$,
\beq
{\rm Tr} [f, A \rho]A^{\dagger},
\label{70}
\eeq
where for simplicity we have dropped the index
of $A^r$, and we observe that it is the limit
when $\varepsilon\to0$ of the expression
\beq
\frac1\varepsilon{\rm Tr}\{e^{\varepsilon f} A\rho\,
e^{-\varepsilon f}A^{\dagger} - A \rho A^{\dagger}\}.
\label{72}
\eeq
The expression (\ref{70}) is 
convex if the first term of (\ref{72}) is convex
as the second term is linear in $\rho$. 
To see that this is the case it suffices to write it in
the following symmetric form
\beq
{\rm Tr}\{(e^{\varepsilon f/2} A\rho^{1/2}
e^{-\varepsilon f/2})(e^{-\varepsilon f/2}\rho^{1/2}
A^{\dagger}e^{\varepsilon f/2}\}.
\eeq

In the general case, instead of terms of the type
(\ref{70}), we face terms of the type 
\beq
{\rm Tr} [f, A^r \rho]B^{r\dagger},
\eeq
which is equal to 
\beq
\sum_s \gamma_{rs}^*{\rm Tr} [f, A^r \rho]A^{s\dagger}.
\label{78}
\eeq
We then expand $A^r$ in another basis set
$A^{\prime s}$ in such a way that the expression
(\ref{78}) becomes diagonal in $A^{\prime s}$, that is,
\beq
\sum_s \lambda_s {\rm Tr} [f, A^{\prime s} \rho]
A^{\prime s\dagger}.
\eeq
Now each of the terms of the summation is similar to 
(\ref{70}) and we may apply the same reasoning above
to show that it is convex. This concludes the
demonstration that $\Pi_r\geq0$ and $\Pi\geq0$.

\section{Fokker-Planck-Kramers structure}

\subsection{Fluctuation and dissipation}

The classical Fokker-Planck-Kramers (FPK) equation
describes a massive particle under the action of a
conservative force plus, a dissipative force, and a
fluctuating force \cite{vankampen1981,tome2015}.
These forces give rise to three terms in the equation
which are the Hamiltonian part, the dissipative part,
and the fluctuation part. In the following we show
that the central equation has a similar structure and
can thus be understood as the quantum version of the
FPK equation.

From the expressions (\ref{69}) and (\ref{24c}),
the central equation can be written in the form
\beq
{i\mu} \frac{d\rho}{dt}
= [H,\rho] - \sum_r [A^r,J^{r\dagger}],
\label{81}
\eeq
where $J^r$ is the current, given by
\beq
J^r = i\mu (B^r \rho -\rho C^{r\dagger}).
\eeq
Defining $D^r = B^{r\dagger} - C^r$, the current becomes
\beq
J^r = i\mu (D^{r\dagger}\rho - [\rho, C^{r\dagger}]).
\eeq
Replacing it in (\ref{81}), the central equation,
it acquires the form
\beq
{i\mu} \frac{d\rho}{dt} = [H,\rho]
+ i\mu \sum_r [A^r, \rho D^{r\dagger} ]
- i\mu \sum_r [A^r,[C^r,\rho]],
\label{87}
\eeq
which has the form of the FPK equation. The three terms
on the right-hand side are respectively, the Hamiltonian
part, the dissipative part, and the fluctuation part. 

Using the relation (\ref{54}), we find 
the following form for the term related to the
dissipation,
\beq
D^r = (\rho^r)^{-1} C^r \rho^r - C^r.
\eeq
Therefore, the equation (\ref{81}) can be
set up if we are given $C^r$ and $\rho^r$.

\subsection{Detailed balance}

The stationary solution $\rho^{\rm st}$ of the central
equation (\ref{81}) is determined by  
\beq
[H,\rho^{\rm st}] - \sum_r[A^r,J^{r\dagger}
(\rho^{\rm st})] = 0.
\label{71}
\eeq
Depending on the covariances of the independent random
variables, that is, on $B^r$ and $C^r$, the stationary
solution may in addition obey a detailed balance condition.
Denoting in this case the stationary solution by
$\rho^{\rm e}$, the detailed balance condition is equivalent
to say that each term of (\ref{71}) vanishes, that is,
$[H,\rho^{\rm e}]=0$ and $J^{r\dagger}(\rho^{\rm e})=0$, or
\beq
B^r \rho^{\rm e} = \rho^{\rm e} C^{r\dagger}.
\label{41b}
\eeq
{\it for all} $r$,
which is the expression of the detailed balance condition. 
This solution is understood as describing the thermodynamic
equilibrium, understood as the state devoid of currents.

We remark that the detailed balance condition (\ref{41b})
should not be confused with the relation (\ref{54})
between $B^r$ and $C^{r\dagger}$. If we use relation
(\ref{54}), the detailed balance condition is just
$\rho^{\rm e}=\rho^r$, for all $r$. Bearing in mind
that the expression of $\rho^r$ for distinct $r$ has
the same form given by equation (\ref{63}), differing
only on the parameter $T_r$, we see that the detailed
balance condition occurs if all the parameters $T_r=T$
are equal. In this case we may identify the common
value $T$ as the temperature of the system and the
$\rho^{\rm e}$ is given by $\beta H=f(\rho^{\rm e})+c$,
or
\beq
\rho^{\rm e} = f^{-1}(\beta H -c),
\eeq
where $f^{-1}$ is the function inverse of $f$.
We recall that the function $f$ is related to the
entropy by $S=\kappa {\rm Tr}f(\rho)$.

If $f=-\ln x$, in which case the entropy is given by
$S=-\kappa {\rm Tr}\rho\ln\rho$,
then $f^{-1}=e^{-x}$ and
\beq
\rho^{\rm e} = \frac1Ze^{-\beta H} ,
\eeq
where $\beta=1/\kappa T$, which is the Gibbs
equilibrium state for a system at a temperature $T$.

\subsection{Non-equilibrium stationary states}

Writing $L^r$ in terms of the current $J^r$, the
expression for the heat flux $\Phi_r$, the entropy flux
$\Psi_r=\Phi_r/T_r$, and the rate of entropy $\Pi_r$
are given by
\beq
\Phi_r = \frac1{i\mu} {\rm Tr}J^{r\dagger}[A^r,H],
\eeq
\beq
\Psi_r
= \frac{\kappa}{i\mu} {\rm Tr}J^{r\dagger}[A^r,f(\rho^r)],
\eeq
\beq
\Pi_r = \frac{\kappa}{i\mu}{\rm Tr}J^{r\dagger}
[A^r, f(\rho)-f(\rho^r)].
\eeq
The total heat flux, the total entropy flux and the
total entropy rate are
are 
\beq
\Phi = \sum_r \Phi_r,
\eeq
\beq
\Psi = \sum_r \Psi_r = \sum_r \frac{1}{T_r}\Phi_r,
\eeq
\beq
\Pi = \sum_r \Pi_r.
\eeq

In the stationary state $\Phi$ and $\Pi+\Psi$ vanish.
If in addition the thermodynamic equilibrium is
established, which occurs if all $T_r$ are equal,
then all currents $J^r$ vanish. In this case $\Phi_r$,
$\Psi_r$, and $\Pi_r$  as well as $\Pi$ and $\Psi$
also vanish. If the stationary state is not an
equilibrium state, which occurs if at least of
$T_r$ is distinct from the others, then at least
one of the currents $J^r$ are nonzero. In this
case, $\Pi$ and $\Psi$ are nonzero, although their
sum vanishes. As $\Pi\geq0$, then in the 
nonequilibrium stationary state
\beq
\Pi=-\Psi>0.
\eeq

\subsection{Examples}

We have proposed a quantum FPK equations which
was obtained through the canonical quantization
\cite{oliveira2016}. For a system of quantum
particles of mass $m$ the quantum  FPK equation
reads \cite{oliveira2016}
\[
i\hbar \frac{\partial\rho}{\partial t} =
[H,\rho] + \frac12\sum_j\gamma_j  [x_j,
\rho g_j + g_j^\dagger \rho]
\]
\beq
+ \frac{\gamma_j m}{i\hbar\beta_j}\sum_j [x_j,[x_j,\rho]],
\eeq
where $x_j$ represents the position of particle $j$,
and
\[
g_j = - \frac{m}{i\hbar\beta_j}
(e^{\beta_j H}x_je^{-\beta_j H} - x_j).
\]
We see that this equation has the form of equation
(\ref{54}) if we set $\mu=\hbar$, $A_j=x_j$, 
\beq
C_j = \frac{\gamma_j m}{\hbar^2\beta_j} x_j,
\quad{\rm and}\quad
D_j = \frac{\gamma_j g_j}{2i\hbar}.
\eeq
The expansion of $g_j$ gives
\beq
g_j = p_j + \frac{\beta_j}{2!}[H,p_j] 
+ \frac{\beta_j^2}{3!}[H,[H,p_j]] + \ldots
\eeq
where $p_j$ is the momentum of particle $j$.
As $g_j$ is related to dissipation we see that
the quantum friction is not generally proportional
to the momentum as is the case of the classical case,
as is manifest in the classical FKP equation.

For bosons the canonical quantization leads to the
following equation \cite{oliveira2018,oliveira2023}
\[
i\hbar \frac{\partial\rho}{\partial t} = [H,\rho]
+ i\gamma_j\sum_j\{[a_j, \rho g_j^\dagger]
-[a_j^\dagger,g_j \rho]\}
\]
\beq
- \frac{i\gamma_j}{\beta_j} \sum_j
\{[a_j,[a_j^\dagger,\rho]]
+ [a_j^\dagger,[a_j,\rho]]\},
\eeq
where $a_j$ and $a_j^\dagger$ represent, respectively,
the annihilation and creation of a boson in a
one-particle state labeled by the index $j$, and 
\beq
g_j = \frac{1}{\beta_j}
(e^{-\beta_j H_j} a_j e^{\beta_j H_j} - a_j).
\eeq
Again we see that this equation has the form of equation
(\ref{54}), if we set $\mu=\hbar$, $A_j=a_j$,
\[
C_j = \frac{\gamma_j}{\hbar\beta_j}a_j,
\quad{\rm and}\quad
D_j = \frac{\gamma_j}{\hbar}g_j.
\]

\section{Conclusion}

We have derived the equations of quantum mechanics
and quantum thermodynamics from the assumption that
a quantum system can be described by an underlying
classical system of particles. Each component 
$\phi_j$ of the wave vector $\phi$is understood as
a complex variable whose real and imaginary parts
are proportional to the coordinate and momentum
associated to a degree of freedom of the underlying
classical system. The equation of motion is
considered to be a stochastic equation so that
$\phi_j$ is a stochastic variable. This result
leads us to conclude that the density matrix $\rho$
obeying either the quantum Liouville or the
Lindblad equation is the covariance matrix 
associated to the random wave vector $\phi$. 
In this sense the present approach gives 
a meaning to the off-diagonal terms of $\rho$.

The understanding of the wave vector $\phi$ as a
stochastic variable and $\rho$ as its covariance
matrix allows an interpretation of quantum
mechanics other than the standard interpretation
\cite{omnes1994,auletta2001,freire2022}.
As the trajectory in the Hilbert space are
stochastic the present approach may fit the
consistent history interpretation of quantum
mechanics \cite{griffiths2002} since the
several trajectories are possible, each 
one occurring with a certain probability.
The present approach is also in accordance
with the standard interpretation if we bear 
in mind that ${\rm Tr}\rho=1$, which means
that $\rho_{ii}$ can be interpreted as a
probability.

\section*{Acknowledgment}

I wish to thank Jacob Barandes for calling my attention
to the papers of Strocchi and Heslot.


\end{document}